# A Practical Route to Demonstrating Core Reactivity in Cs Polyoxides under Pressure

Yuanhui Sun[1], Samantha Ellis[1], Dalar Khodagholian[1], Lansen Wyke[1], Austin Ellis[1], Peter C. Müller[2], Boris N. Schüpp[2], Ryky Nelson[2], Richard Dronskowski[2,3], Maosheng Miao[1,*]

[1]Department of Chemistry and Biochemistry, California State University Northridge, CA 91303, USA

[2]Chair of Solid-State and Quantum Chemistry, Institute of Inorganic Chemistry, RWTH Aachen University, 52056 Aachen, Germany

[3]Hoffmann Institute of Advanced Materials, Shenzhen Polytechnic, Shenzhen, China

*Email: mmiao@csun.edu*

**Abstract**

Activating normally inert core electrons expands the bonding possibilities of elements beyond conventional valence chemistry. Although Cs core reactivity under pressure was predicted in polyfluorides more than a decade ago, experimental confirmation remains elusive, largely because loading highly corrosive $F_2$ poses serious challenges for diamond anvil cells. Here, global crystal structure searches combined with first-principles calculations identify Cs polyoxides as a practical alternative for investigating core reactivity. Compression activates Cs 5p electrons, leading to oxidation beyond the usual +1 state and a striking reversal of conventional polyoxide behavior: O-O bonds disappear, while strong Cs-O covalent bonds generate molecular units and extended sheets. Most importantly, CsO, a solid peroxide stable at ambient pressure, is predicted to undergo this transformation at 221 GPa, forming linear $[CsO_2]^-$ units. This approach replaces a chemical reaction between separate reactants with a phase transition in a single solid, avoiding $F_2$ loading and simplifying experimental implementation without changing the overall composition. The loss of peroxide O-O pairs provides a vibrational signature of the transformation, while calculated infrared and Cs $L_3$-edge XANES spectra, together with simulated X-ray diffraction patterns, guide experimental identification. These results establish an experimentally testable route toward core reactivity and show how pressure can extend the chemical possibilities of a familiar alkali metal.

## Introduction

High pressure can stabilize compounds with unconventional compositions and containing various homonuclear species such as polycarbon,[1] polynitrogen,[2] polyoxygen[3–5] and polyhalogen[6] anions, by enhancing the strength of homonuclear bonds.[7] Therefore, it is a general trend that more polyoxides, such as peroxides,[3,4] superoxides[8,9] and ozonides[5] that contain stronger O–O bonds become stable under high pressure. On the other hand, high pressure can also change the composition of stable compounds by altering the relative energies of different atomic orbitals.[7,10] The most striking phenomenon caused by this effect is the activation of the core electrons in forming new bonds.[11] It is not yet known what will happen if both effects are strong in the same material.

Despite the strong aspiration, all the previous attempts of achieving core reactivity under ambient pressure have failed due to the ultrahigh stability of the core shells in all elements.[12,13] The prediction and synthesis of $HgF_4$ attained unusual oxidation state of +4 for Hg and demonstrated the reactivity of 5*d* semi-core electrons.[14] The obtained $HgF_4$ molecule is not stable and show extremely short lifetime in cryogenic He matrix. In stark contrast, high pressure can stabilize Hg in high oxidations states and form thermodynamically stable $HgF_3$ and $HgF_4$ compounds.[15] More strikingly, as shown by density functional theory (DFT) calculations, Cs can react with excess amount of $F_2$ under pressure and form stable $CsF_n$ ($n$>1) compounds consisting of Cs in oxidations states from +2 to +6, which can only occur while its 5*p* core electrons are activated and form bonds with F.[11,16]

Regardless of the DFT predictions of $CsF_n$ ($n$>1) compounds at pressures that can be readily accessed by the current high-pressure methods, the inevitable damage to the diamond anvil cell (DAC) by loading the highly corrosive $F_2$ liquid obstruct its experimental demonstrations, and the core reactivity of any kind remains as a theory. Among many possible methods to overcome this difficulty, the load of solid, less corrosive, and yet sufficiently strong oxidants is most practical, and the search of such oxidants is an unaccomplished mission of computational high-pressure chemistry. Aiming at this goal, we conducted a full-scale study of Cs polyoxides under pressure, including thorough crystal structure searches of a series $Cs_2O_n$ ($n$=1, 2, …, 7) compounds in a pressure range from 0 to 500 GPa.

## Computational methods

**Structure search.** The search for the structures of Cs-O compounds was carried out based on a global minimum search of the free energy landscape with respect to structural variations by combining the particle swarm optimization (PSO) algorithm with first-principles energetic calculations.[17,18] To find the structures that realize the high oxidation state of Cs atoms, we performed global structure searches on O-rich compounds with various $Cs_2O_n$ ($n$ = 1 - 7) compositions at 0 K and the selected pressures of 100, 200, 300, 400, and 500 GPa, respectively. In addition, the known O-rich Cs-O compounds are included and their energies are compared with structure search results.[8,19–21] By evaluating the total energy of these structures, 60% of them with lowest enthalpies together with 40% newly generated structures are regarded as the next generation of structures by the structure operators of PSO. The procedure stopped after 20 generations were created. Numerous works using this code show high accuracy and efficiency.[22–24]

**Electronic structure calculations.** We conducted calculations on the formation enthalpies and electronic properties using density-functional-theory (DFT) as implemented in Vienna *ab initio* simulation package (VASP) code.[25] The DFT calculations were performed using a generalized gradient approximation[26] exchange-correlation functional within the framework of Perdew-Burke-Ernzerhof density functional.[27] The projector augmented wave[28] (PAW) method was used as a pseudopotential approach with $5s^2 5p^6 6s^1$ and $2s^2 2p^4$ valence electrons for Cs and O atoms respectively. We adopted a kinetic energy cutoff of 520 eV for wave-function expansion and a $k$-point mesh of $2\pi \times 0.03$ Å$^{-1}$ or less for Brillouin zone integration. The structures (including lattice parameters and atomic positions) were fully optimized until the residual forces were converged within 0.02 eV/Å. The dynamical stability of predicted structures was determined by phonon calculations using the finite displacement approach[29] as implemented in the Phonopy code.[30] Crystal orbital Hamilton population (COHP)[31] and crystal orbital bond index (COBI)[32] analyses giving information on the interatomic interaction are implemented in the LOBSTER package.[33–36] To account for potentially unusual oxidation states and orbital occupations not necessarily well described by tabulated STO basis sets, numerical basis sets directly derived from the PAW pseudopotentials were used.

**Formation enthalpy calculations.** Formation enthalpy was used to learn the stability of Cs-O compounds, which is calculated by the following formula:

$$h_f(\mathrm{Cs_2O_n}) = \left[H(\mathrm{Cs_2O_n}) - H(Cs_2O) - \frac{(n-1)H(\mathrm{O_2})}{2}\right]/(n+2) \quad (1)$$

The formation enthalpy is the differences between sum of enthalpies of products and enthalpies of reactants in reaction $\mathrm{Cs_2O} + \frac{n-1}{2}\mathrm{O_2} \rightarrow \mathrm{Cs_2O_n}$. Thanks to the exceeding stability of $Cs_2O$ throughout the studied pressure range, using $Cs_2O$ as a reactant and the above reaction will not change the convexity, i.e. a compound that is shown stable/unstable using the reaction enthalpy as defined above will also be shown stable/unstable if the enthalpy of formation, which is the enthalpy change of the reaction $2\mathrm{Cs} + \frac{n}{2}\mathrm{O_2} \rightarrow \mathrm{Cs_2O}_n$, is used to construct a convex hull. The predicted structure with the lowest enthalpy for each composition is used to evaluate the formation enthalpy with respect to the most stable $Cs_2O$ and $O_2$ molecules. $h_f$ is the formation enthalpy per atom. For the solid lines in Figure 1a, the compounds located on the convex hull are stable against decomposition, because decomposing into other compounds will cost energy.[37]

## Results and discussion

**Stability.** Cs-O compounds show distinctly different thermodynamic behaviors below and above 300 GPa (Figure 1). Several $Cs_nO_m$ compounds such as CsO, $Cs_2O_3$ and $CsO_2$ are stable under ambient pressure (Figure S1). CsO remains stable up to 500 GPa, the highest pressure studied in this work. Remarkably, $Cs_2O_3$ becomes unstable at 11.3 GPa but reappears as a stable compound in the pressure range from 346 GPa to 488 GPa. $CsO_2$ becomes unstable at 346 GPa, whereas a new compound with even higher O composition, $CsO_3$, becomes stable at 306 GPa. All the compounds are found to be dynamically stable in their stable pressure ranges (Figure S2).

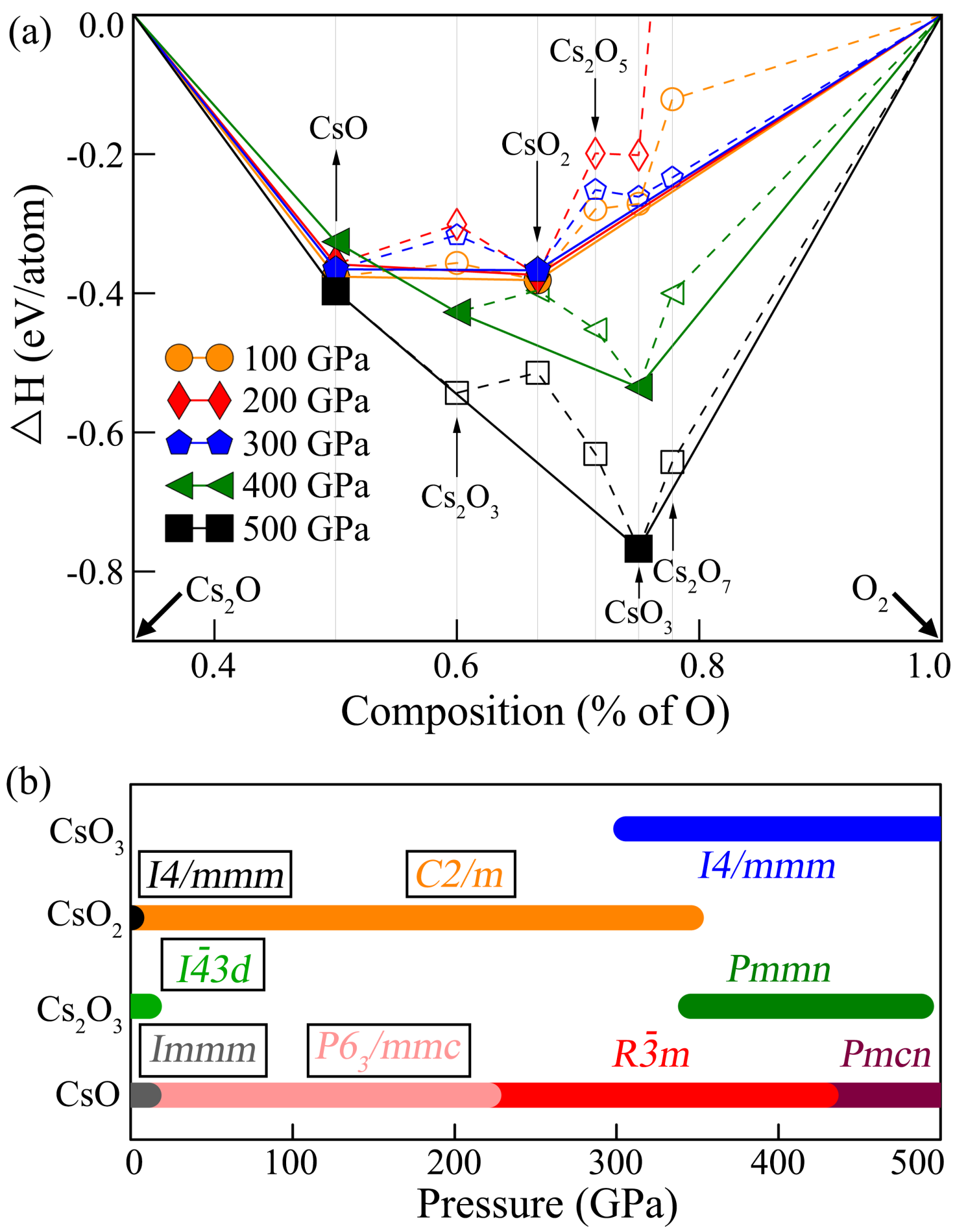

**Figure 1.** Stability of Cs-O compounds under high pressure. (a) Enthalpies of formation of $Cs_2O_n$ under 100, 200, 300, 400, and 500 GPa, respectively. Solid lines denote the convex hull and dashed lines connect data points. (b) Pressure-composition phase diagram of Cs-O compounds within the estimated pressure range from 0 to 500 GPa.

**Crystal structures.** The most important structural change to Cs-O compounds under high pressure is the disappearance of polyoxygen anions that are the key structural component of polyoxides. At 11 GPa, CsO transforms from *Immm* structure[38] into a $P6_3/mmc$ structure in which Cs atoms form a double hexagonal close pack (dhcp) lattice and the $O_2^{2-}$ anions occupy one set of the octahedral sites (Figure 2a). The O–O bond length in CsO is 1.432 Å at 100 GPa, which is close to the O–O bond length of 1.460 Å in $H_2O_2$.[9] It further transforms into a $R\bar{3}m$ structure (Figure 2b) at 221 GPa and an orthogonal structure with *Pmcn* symmetry (Figure 2c) at 429 GPa, respectively. The $R\bar{3}m$ structure can be derived from a CsCl structure by slightly compressing along (111) direction. Both structures show no O-O bond as in typical peroxides, and the minimum O-O distances are 2.538 Å in $R\bar{3}m$ structure at 300 GPa and 2.575 Å in *Pmcn* structure at 500 GPa, respectively. Most strikingly, $R\bar{3}m$ CsO consists of $CsO_2$ molecules. Half of its Cs atoms ($Cs^1$) possesses two 1st nearest and six 2nd nearest O atoms, and the corresponding Cs–O distances are 1.940 and 2.494 Å at 221 GPa. The other Cs atoms ($Cs^2$) possess two 1st nearest O atoms with a distance of 2.211 Å and six 2nd nearest O atoms with a distance of 2.418 Å. *Pmcn* CsO shows no molecular features and all its Cs (O) atoms have six nearest neighbor O (Cs) atoms, with Cs–O distances ranging from 2.092 to 2.175 Å.

The low-pressure structure of $Cs_2O_3$ ($I\bar{4}3d$)[21] contains O–O bonds, but the high-pressure structure (*Pmmn*, Figure 2d) does not. There are two types of Cs atoms in *Pmmn* structure, at 400 GPa one ($Cs^1$) shows slightly shorter Cs–O distances of 1.953, 2.017, and 2.087 Å with 4, 2, and 2 neighboring O atoms, and the other ($Cs^2$) shows Cs-O distances of 2.112, 2.179, and 2.249 Å with 1, 2, and 4 O atoms. $Cs^1$ and O atoms form a buckled two-dimensional sheet. The Cs superoxide ($CsO_2$) exhibits O–O bonds throughout its stable pressure range from 0 to 346 GPa. It transforms from the *I*4/*mmm* structure[38] at ambient pressure to a *C*2/*m* structure (Figure 2e) at 0.5 GPa and shows no further structure change until $CsO_2$ becomes unstable at 346 GPa. The O-O bond length is 1.351 Å in *I*4/*mmm* under ambient pressure and 1.278 Å in *C*2/*m* structure under 200 GPa. $CsO_3$ adopts an *I*4/*mmm* structure that contains no O–O bonds (Figure 2f). Instead, its O atoms locate at both tetrahedral centers and square centers of Cs atoms. Correspondingly, the coordination number of Cs is 12 that is one of the highest among all inorganic compounds.

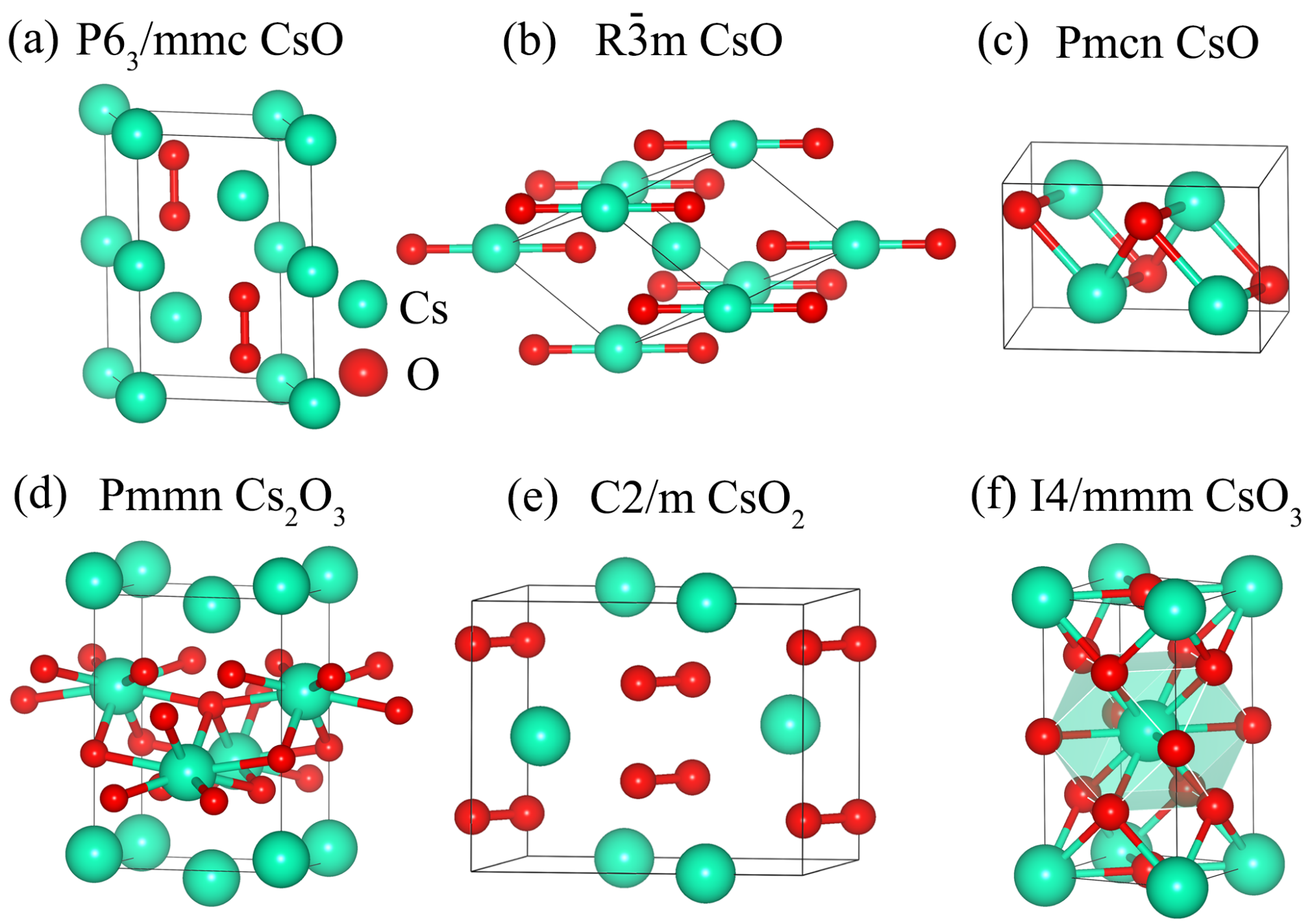

**Figure 2.** Crystal structures of Cs-O compounds. (a) CsO in $P6_3/mmc$ symmetry at 100 GPa. (b) CsO in $R\bar{3}m$ symmetry at 300 GPa. (c) CsO in *Pmcn* symmetry at 500 GPa. (d) $Cs_2O_3$ in *Pmmn* symmetry at 400 GPa. (e) $CsO_2$ in *C2/m* symmetry at 200 GPa. (f) $CsO_3$ in *I4/mmm* symmetry at 400 GPa. The cesium and oxygen atoms are colored spring green and red, respectively.

**Electronic structures and the bonding features.** The abnormal disappearance of O–O bonds in Cs polyoxides under high pressure implies a substantial change of the electronic states in these compounds. Both the analysis of the electron states and the calculations of the total charge transfers unveil that it is caused by the activation of Cs 5*sp* electrons. The projected density of states (PDOS) shows that the Cs 5*p* states in a typical peroxide, such as CsO in $P6_3/mmc$ structure, already extend and largely overlap with O 2*p* states up to the Fermi level under high pressure (Figure 3a). However, COHP/COBI calculations[31] show no bonding between Cs 5*p* and O 2*p* (Figure S4a and S4b), and the total charges calculated in Bader's Quantum Theory of Atoms in Molecules (QTAIM)[39] show a typical +1 oxidation state for Cs (Figure 4). On the other hand, COHP calculations show strong covalent O–O bonds in $P6_3/mmc$ CsO with an integrated COHP (ICOHP) value of −8.426 eV/pair (Table S2) as well as an integrated COBI of 0.826 per bond which is in good agreement with the expected single-bond order of the $O_2^-$ ion. In contrast, $R\bar{3}m$ CsO shows strikingly different chemistry, although the compound seems just undergo a structure transformation. Two features of its PDOS (Figure 3b), namely reduced gap (also Figure S3) and the large number of 5*p* states above the Fermi level, reveal the electron transfer from Cs 5*p* to O 2*p* states, which is consistent with the Bader charges that clearly show an oxidation state beyond +1 for $Cs^1$ atoms. Both Electron Localization Function (ELF)[40] and COHP/COBI calculations show strong covalent Cs–O bonds within $(CsO_2)^-$ anionic molecules (Figure 5 and Figure S4c and S4d). The ICOHP value of $Cs^1$–O bond is -3.097 eV/pair at 221 GPa which is distinctly significant than the ICOHP value of −0.719 eV/pair for the $Cs^2$–O bonds. For comparison, the ICOHP value of Cs–O in $P6_3/mmc$ CsO is −0.806 eV/pair at the same pressure. The *Pmcn* CsO is metallic and the states at the Fermi level show large Cs 5*p* components, and the Bader charges also show an oxidation state of Cs beyond +1. The ICOHP values for the shortest Cs–O bond is −1.030 eV/pair at 500 GPa (Table S2).

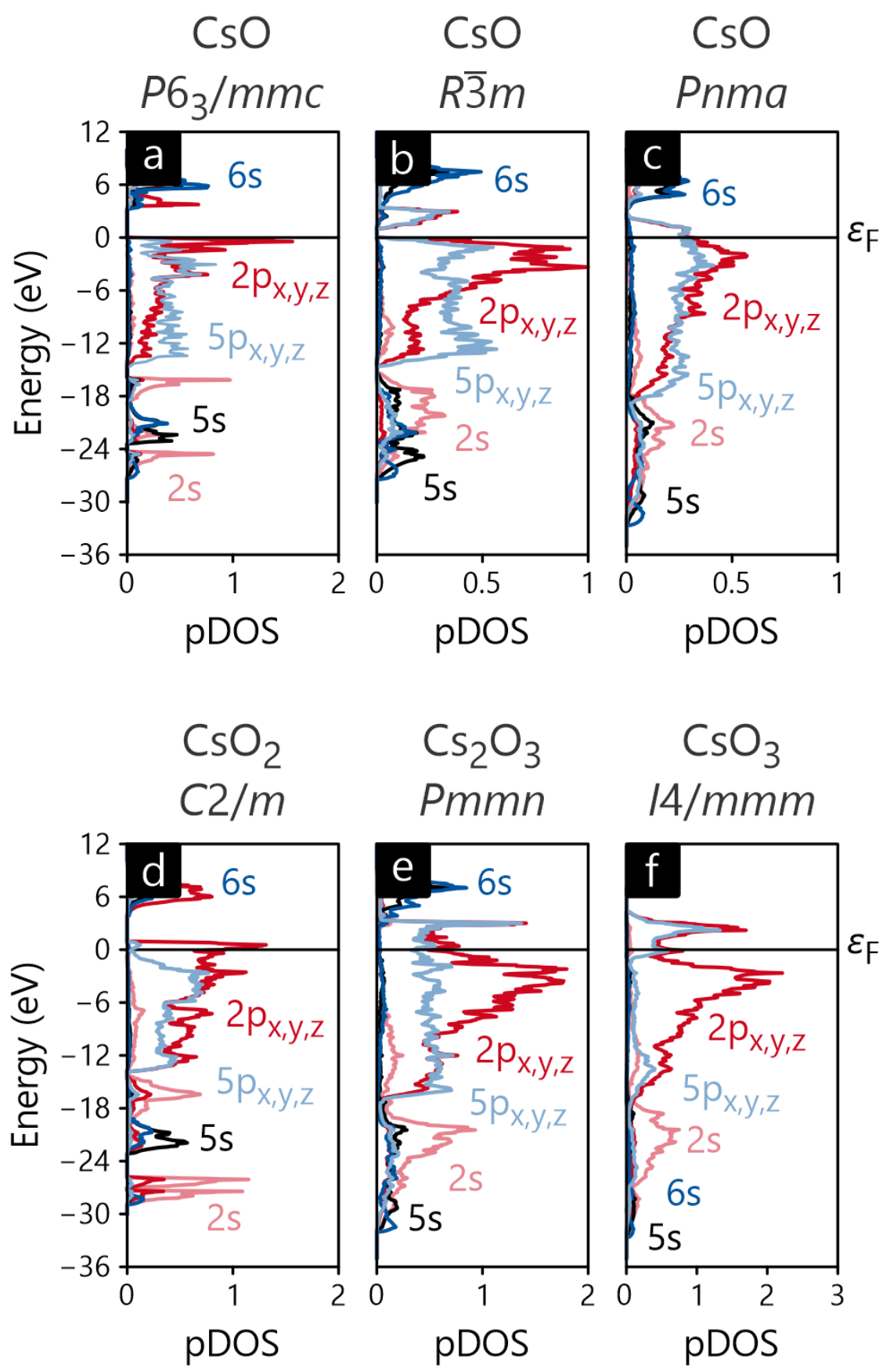

**Figure 3.** Electronic structures of Cs-O compounds. Calculated PDOSs for (a) CsO in $P6_3/mmc$ symmetry at 221 GPa, (b) CsO in $R\bar{3}m$ symmetry at 221 GPa, (c) CsO in *Pmcn* symmetry at 500 GPa, (d) $CsO_2$ in *C2/m* symmetry at 200 GPa, (e) $Cs_2O_3$ in *Pmmn* symmetry at 400 GPa, and (f) $CsO_3$ in *I4/mmm* symmetry at 400 GPa, respectively.

Like *Pmcn* CsO, *C*2/*m* $CsO_2$ and *Pmmn* $Cs_2O_3$ are also metallic. However, the states of $CsO_2$ at the Fermi level show only minor contributions from Cs 5*p*, which is consistent with the low Bader charge of 0.88 *e* (Figure 4a). The low ICOHP values, −0.437 eV/pair, reveal a very weak Cs–O bond, agreeing with a typical superoxide that consists of $Cs^+$ cations and $(O_2)^-$ anions. On the other hand, the Bader charges Cs atoms in $Cs_2O_3$ reveal a striking multivalent nature of Cs ions. While $Cs^2$ shows a charge of 0.87 at 400 GPa, the Bader charge of $Cs^1$ is as high as 2.17, corresponding to a high oxidation state. $Cs^1$ also forms stronger bonds with neighboring O atoms. The ICOHP value of the shortest $Cs^1$–O bond is −2.376 eV/pair. For comparison, the value of the shortest $Cs^2$–O bond is −0.486 eV/pair (Table S2). Thus, it proves that $Cs_2O_3$ consists of $Cs^1$–O monolayer and isolated $Cs^2$ anions due to the large participation of 5*p* electrons in $Cs^1$. Correspondingly, the chemical formula of $Cs_2O_3$ can be explicitly noted as $Cs^+(CsO_3)^-$. To cross-validate our analysis, the wave-function-based Löwdin charges of Cs–O compounds are calculated as shown in Figure 4b, which is quite similar to the evolutions of Bader charge results.

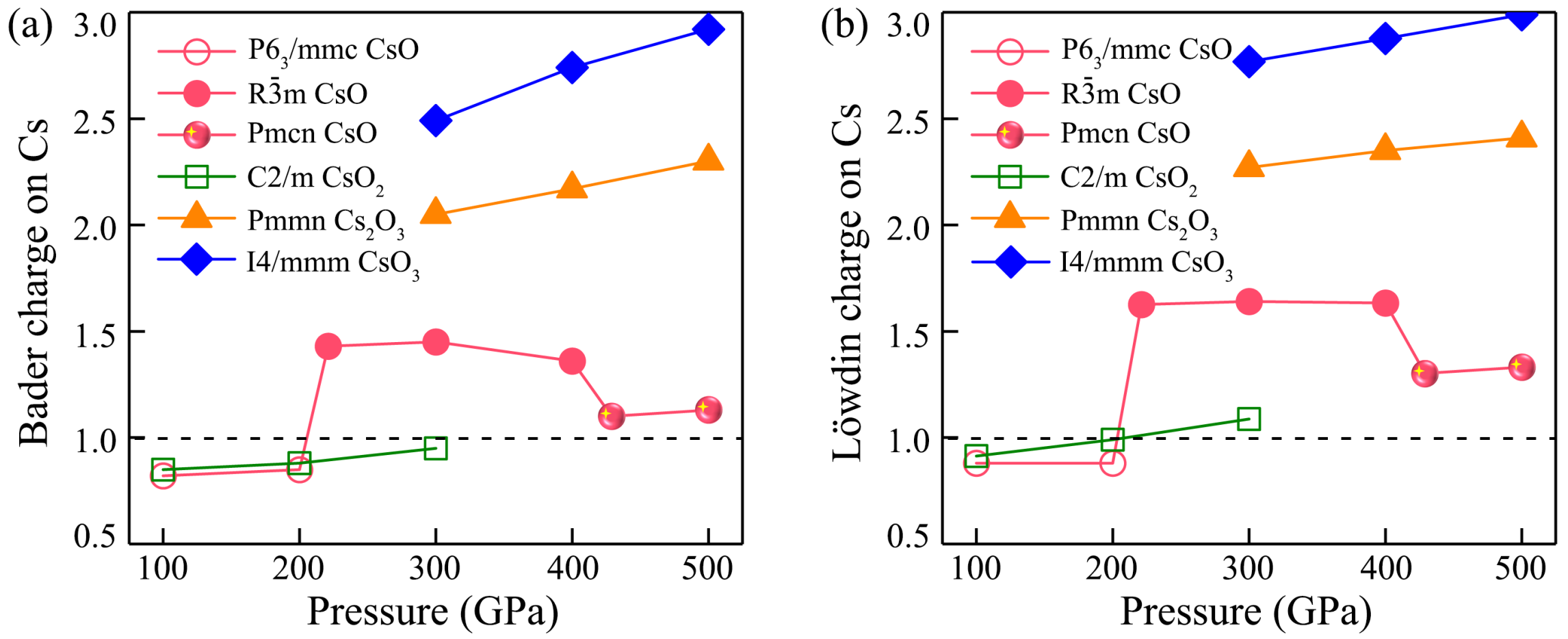

**Figure 4.** Mechanism of pressure-driven high oxidation states. Calculated (a) Bader charge and (b) Löwdin charge of Cs in Cs-O compounds in their stable pressure ranges.

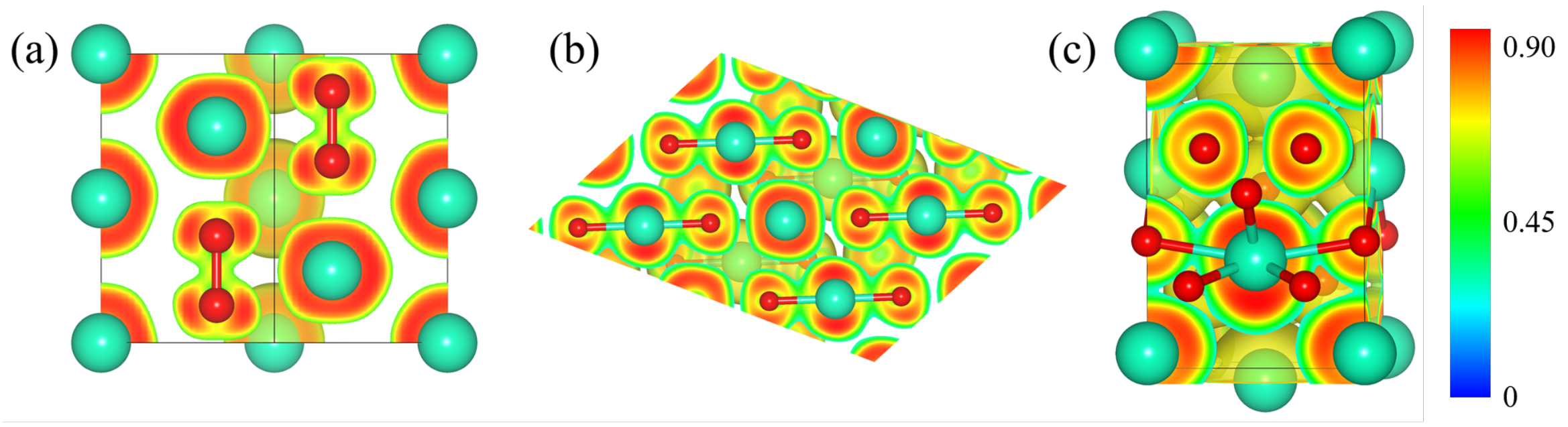

**Figure 5.** Electron localization function of (a) *P6₃/mmc* CsO in the [110] plane at 221 GPa, (b) $R\bar{3}m$ CsO in the $[01\bar{1}]$ plane at 221 GPa, (c) and *Pmmn* $Cs_2O_3$ at 400 GPa, respectively. Isosurface value is 0.5.

**Pressure-driven core reactivity.** The abnormal behavior of Cs polyoxides under pressure is caused by the activation of Cs 5*sp* electrons and their involvement in forming chemical bonds. Pressure can change the relative energies of different atomic orbitals due to their different size and screening effects of the inner-shell orbitals. Most dramatically, it can activate the core electrons if their orbital energies become closer to or even higher than the orbital energies of oxidant atoms, such as F and O. Using a He matrix model, we find that the energy of Cs 5*p* orbital becomes higher than that of O 2*p* orbital at 318 GPa, indicating strong bonding and large electron transfer may occur between the two orbitals (Figure 6). Interestingly, the energy of Ra 5*p* orbital also crosses the energy of O 2*p* orbital at 362 GPa, indicating that Ra polyoxides, such as $RaO_2$,[41] might show the same abnormal behavior in the similar pressure range. As a matter of fact, many alkali and alkaline earth metals polyoxides might undergo the same chemical transformations as CsO and $Cs_2O_3$. However, the required pressures will be much higher and harder to access by current high-pressure approaches.

Additional insight comes from the wavefunction-derived Löwdin charges because, in contrast to Bader charges, they allow to break down the atomic charges into orbital-resolved ones. For example, the 500 GPa CsO phase crystallizing in *Pmcn* shows Löwdin populations of 1.23 (5*s*), 5.42 (5*p*), and 1.09 (6*s*) for Cs, corresponding to orbital charges of +0.77 (5s), +0.58 (5p), and –0.09 (6s) and a total charge of +1.26, marginally higher than the Bader charge of +1.1. Hence, the entire 5*sp* shell is involved, also reflecting some mixing of 5*s* and 6*s* in the bonded atom, despite being orthogonal for the isolated atom.

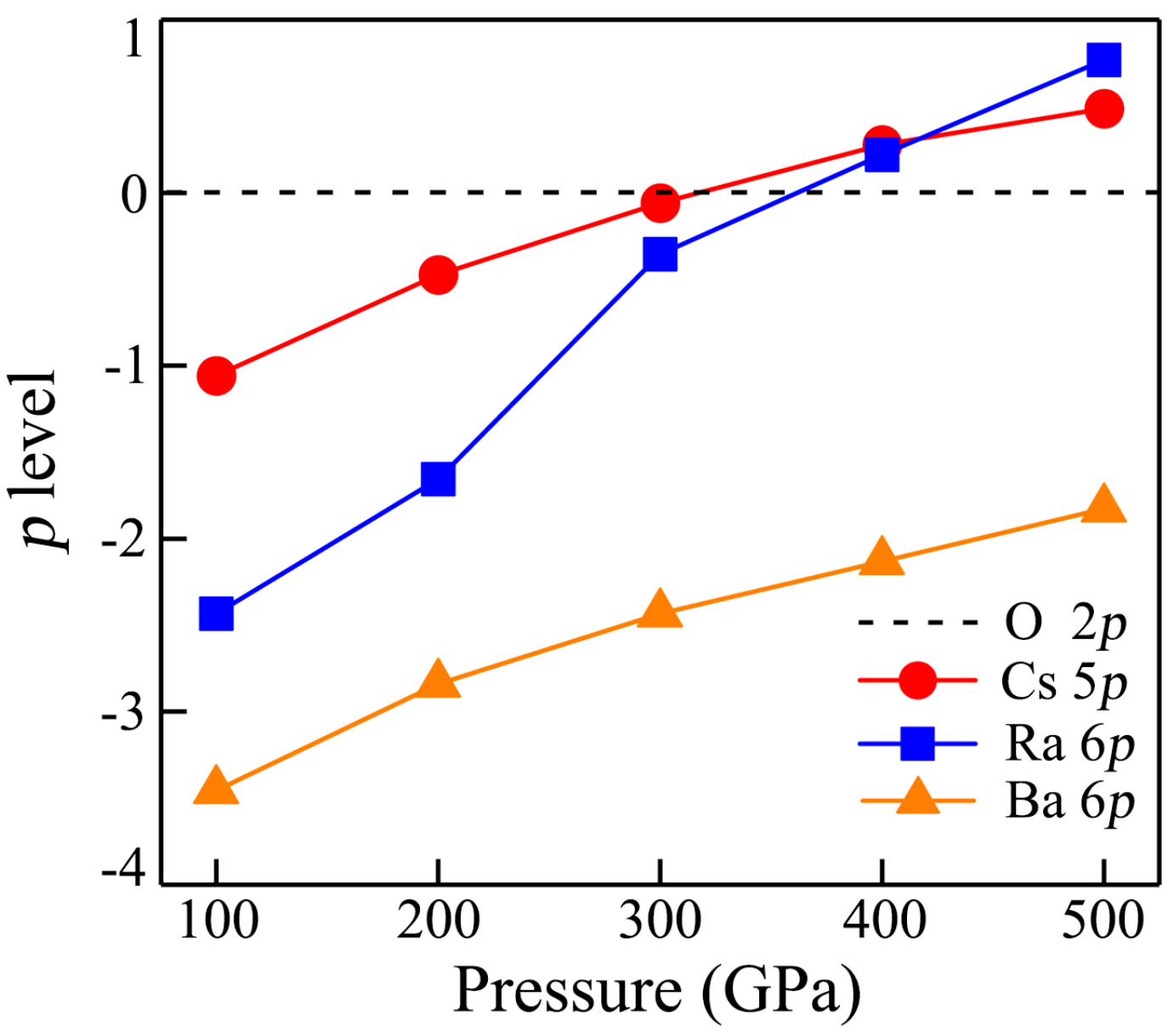

**Figure 6.** Atomic orbital energy levels of the outermost fill *p* levels of selected elements as a function of pressure, including O, Cs, Ra, and Ba atoms. The energies of O atom are set to zero for given pressures. Pressure effect is modeled by putting elements in a face-centered cubic supercell (3×3×3) He matrix, in which one He atom is replaced by the atom being examined.[42]

**Observing core reactivity in a lab.** The abnormal changes of structure and chemistry of Cs polyoxides provide a unique approach to achieve core reactivity in high-pressure experiments. The major advantage of polyoxides is that many of them can be synthesized as solid samples under ambient pressure and are ready for loading into DAC. Furthermore, despite the dramatic chemical change of the compound, the composition remains the same, eluding the need of reacting two substances in DAC. Among all polyoxides, CsO is the most promising candidate. Solid CsO can be obtained under ambient pressure and loaded into DAC. Around 221 GPa, it will transform into the $R\bar{3}m$ structure that contains $CsO_2$ molecules bonded by Cs 5*p* electrons and show a volume reduction (Figure S6). The trade-off of this approach is the requirement of higher pressure around 220 GPa, comparing with the synthesis of Cs polyfluorides under 30 – 50 GPa.[11] Many methods can be used to identify the $R\bar{3}m$ structure, including the direct measurement by XRD (Figure S5) and the measurement of the vibrational modes by IR and Raman spectroscopy. Furthermore, the dramatic change of the band gap can be measured by absorption or emission spectroscopy.

In addition to XRD and the measurement of vibrational modes by IR and Raman spectroscopy, Cs $L_3$-edge XANES can be used to detect the electronic structure changes associated with core reactivity in CsO. We calculated the Cs $L_3$-edge XANES spectra of CsO from ambient pressure to 250 GPa (Figure 7a). The spectra show a systematic evolution of the white-line and near-edge region with increasing pressure, including a shift of the main absorption feature to higher energy and a distinct change in peak shape near the 221 GPa phase. These changes are consistent with the abnormal change of the local Cs coordination and the activation of Cs 5sp electrons under high pressure. The calculated IR spectrum of the 221 GPa phase provides a complementary vibrational fingerprint for identifying this structure in high-pressure experiments (Figure 7b). Together, the XANES and IR spectra provide experimentally accessible signatures of the core reactivity predicted for CsO. Additional calculated Cs $L_3$-edge XANES spectra are provided including site- averaged spectra for Cs–O stoichiometries at 250 GPa and a reference Cs compound (Figure S7).

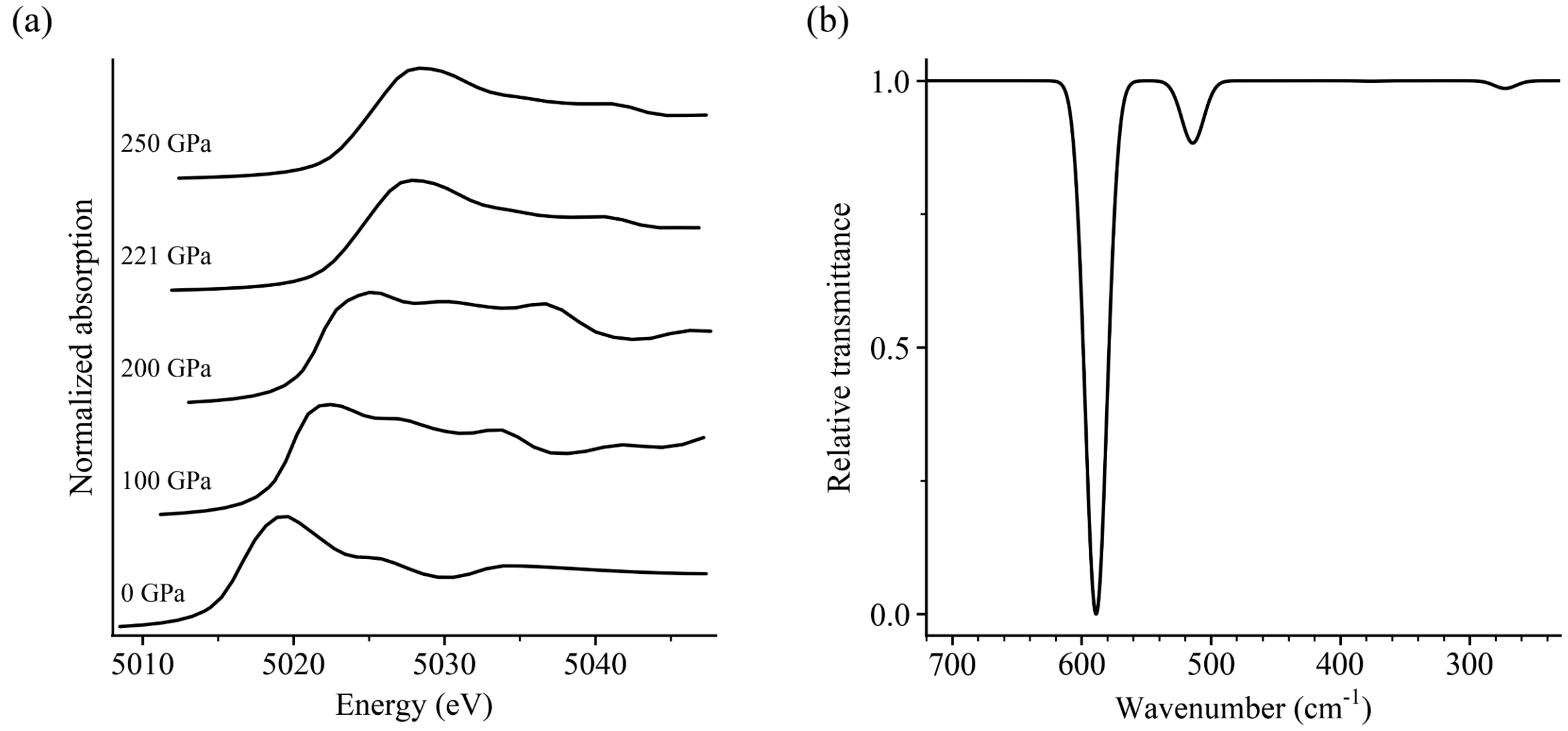

**Figure 7.** Calculated XANES and IR spectra of CsO under high pressure. (a) Cs $L_3$-edge XANES spectra of the stable CsO structures from 0 to 250 GPa. The spectra are normalized and vertically shifted for comparison. No horizontal energy shift was applied. (b) Calculated IR spectrum of CsO at 221 GPa plotted as relative transmittance.

## Conclusion

DFT calculations and PSO crystal structure searches on a series of Cs-O compounds show striking abnormal changes of Cs polyoxides under high pressure, including the disappearance of O-O bonds, the formation of strong Cs-O covalent bonds, the presence of the multivalent Cs ions, and the exhibition of $CsO_2$ linear molecules and $CsO_3$ monolayers. The electronic structure analysis reveal that the abnormality is caused by the activation of the Cs 5*p* (and also 5*s*) electrons under high pressure, which renders Cs a *p*-block element. Many polyoxides can be synthesized under ambient pressure and can be loaded into DAC as solid samples, which provides a new method to observe the core reactivity in high-pressure experiments. Our work shows that CsO is the most promising candidate. It undergoes a structural transition accompanied by the activation of Cs 5*p* electrons under 221 GPa, which causes many atomic and electronic structure changes that can be detected by various methods such as XANES, XRD, IR, Raman etc.

## Associated Content

## Supporting Information

Detailed computational methods; formation enthalpies and phonon spectra; electronic band structures and COHP/COBI analyses; simulated X-ray diffraction patterns, equations of state, and Cs $L_3$-edge XANES spectra; structural parameters and fractional atomic coordinates; and interatomic distances, ICOHPs, and ICOBIs.

**Supporting information for**

# A Practical Route to Demonstrating Core Reactivity in Cs Polyoxides under Pressure

Yuanhui Sun[1], Samantha Ellis[1], Dalar Khodagholian[1], Lansen Wyke[1], Austin Ellis[1], Peter C. Müller[2], Boris N. Schüpp[2], Ryky Nelson[2], Richard Dronskowski[2,3], Maosheng Miao[1,*]

[1]Department of Chemistry and Biochemistry, California State University Northridge, CA 91303, USA

[2]Chair of Solid-State and Quantum Chemistry, Institute of Inorganic Chemistry, RWTH Aachen University, 52056 Aachen, Germany

[3]Hoffmann Institute of Advanced Materials, Shenzhen Polytechnic, Shenzhen, China

**Email: mmiao@csun.edu*

# Table of Contents

## 1. Computational methods in detail

**Structure search.** The search for the structures of Cs-O compounds was carried out based on a global minimum search of the free energy landscape with respect to structural variations by combining the particle swarm optimization (PSO) algorithm with first-principles energetic calculations.[1,2] To find the structures that realize the high oxidation state of Cs atoms, we performed global structure searches on O-rich compounds with various $Cs_2O_n$ ($n$ = 1 - 7) compositions at 0 K and the selected pressures of 100, 200, 300, 400, and 500 GPa, respectively. In addition, the known O-rich Cs-O compounds are included and their energies are compared with structure search results.[3–6] By evaluating the total energy of these structures, 60% of them with lowest enthalpies together with 40% newly generated structures are regarded as the next generation of structures by the structure operators of PSO. The procedure stopped after 20 generations were created. Numerous works using this code show high accuracy and efficiency.[7–9]

**Electronic structure calculations.** We conducted calculations on the formation enthalpies and electronic properties using density-functional-theory (DFT) as implemented in Vienna *ab initio* simulation package (VASP) code.[10] The DFT calculations were performed using a generalized gradient approximation[11] exchange-correlation functional within the framework of Perdew-Burke-Ernzerhof density functional.[12] The projector augmented wave[13] (PAW) method was used as a pseudopotential approach with $5s^2 5p^6 6s^1$ and $2s^2 2p^4$ valence electrons for Cs and O atoms respectively. We adopted a kinetic energy cutoff of 520 eV for wave-function expansion and a $k$-point mesh of $2\pi \times 0.03$ Å$^{-1}$ or less for Brillouin zone integration. The structures (including lattice parameters and atomic positions) were fully optimized until the residual forces were converged within 0.02 eV/Å. The dynamical stability of predicted structures was determined by phonon calculations using the finite displacement approach[14] as implemented in the Phonopy code.[15] Crystal orbital Hamilton population (COHP)[16] and crystal orbital bond index (COBI)[17] analyses giving information on the interatomic interaction are implemented in the LOBSTER package.[18–21] To account for potentially unusual oxidation states and orbital occupations not necessarily well described by tabulated STO basis sets, numerical basis sets directly derived from the PAW pseudopotentials were used.

**Formation enthalpy calculations.** Formation enthalpy was used to learn the stability of Cs-O compounds, which is calculated by the following formula:

$$h_f(\mathrm{Cs_2O_n}) = \left[H(\mathrm{Cs_2O_n}) - H(Cs_2O) - \frac{(n-1)H(\mathrm{O_2})}{2}\right]/(n+2) \quad (1)$$

The formula shows formation enthalpy is equal to the differences between sum of enthalpies of products and enthalpies of reactants in reaction $\mathrm{Cs_2O} + \frac{n-1}{2}\mathrm{O_2} \rightarrow \mathrm{Cs_2O_n}$. Thanks to the exceeding stability of $Cs_2O$ throughout the studied pressure range, using $Cs_2O$ as a reactant and the above reaction will not change the convexity, i.e. a compound that is shown stable/unstable using the reaction enthalpy as defined above will also be shown stable/unstable if the enthalpy of formation, which is the enthalpy change of the reaction $2\mathrm{Cs} + \frac{n}{2}\mathrm{O_2} \rightarrow \mathrm{Cs_2O}_n$, is used to construct a convex hull. The predicted structure with the lowest enthalpy for each composition is used to evaluate the formation enthalpy with respect to the most stable $Cs_2O$ and $O_2$ molecules. $h_f$ is the formation enthalpy per atom. For the solid lines in Figure 1a, the compounds located on the convex hull are stable against decomposition, because decomposing into other compounds will cost energy.[22]

## 2. Thermal and phonon stability of Cs-O compounds

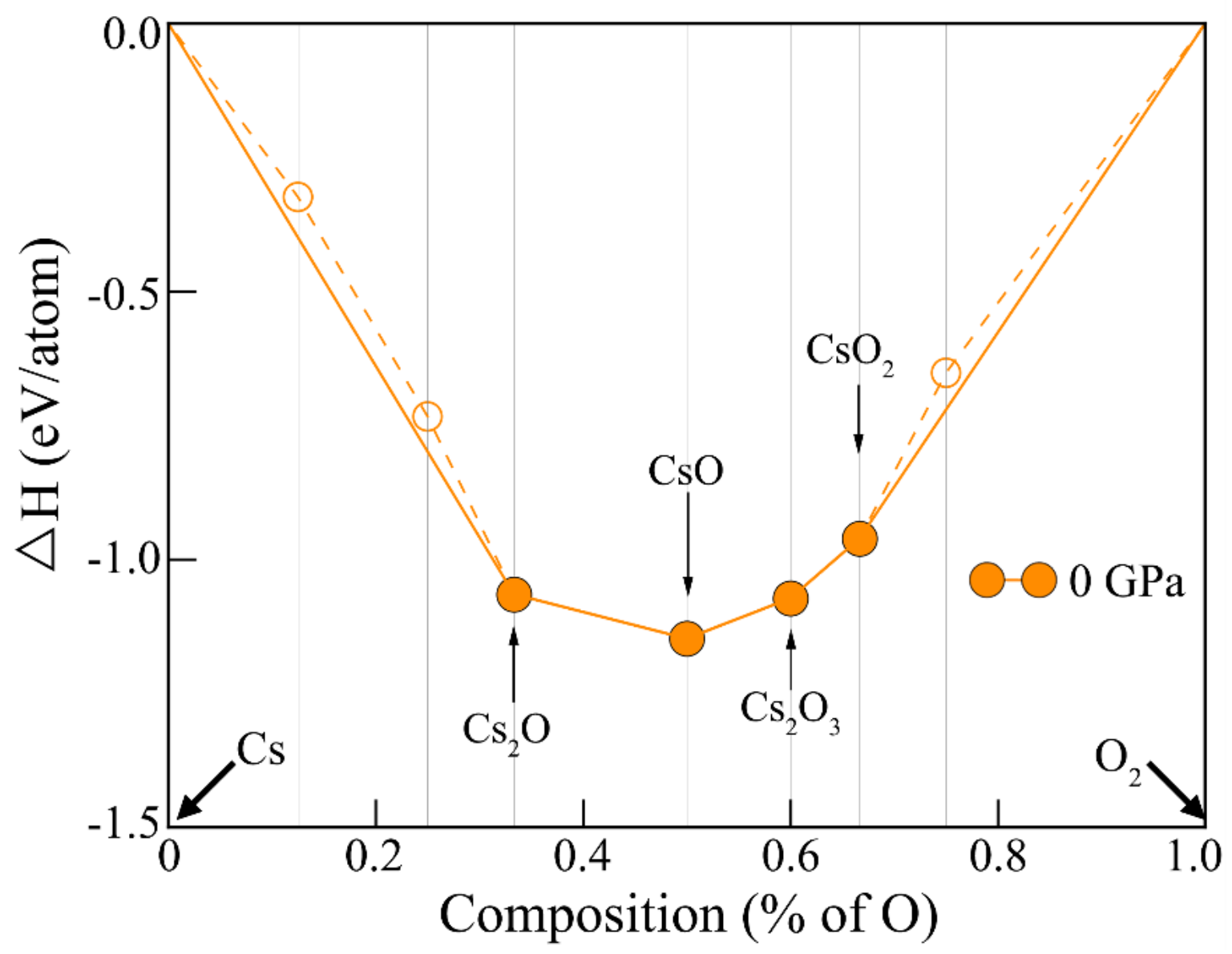

**Figure S1.** Enthalpies of formation of $Cs_2O_n$ at 0 GPa.

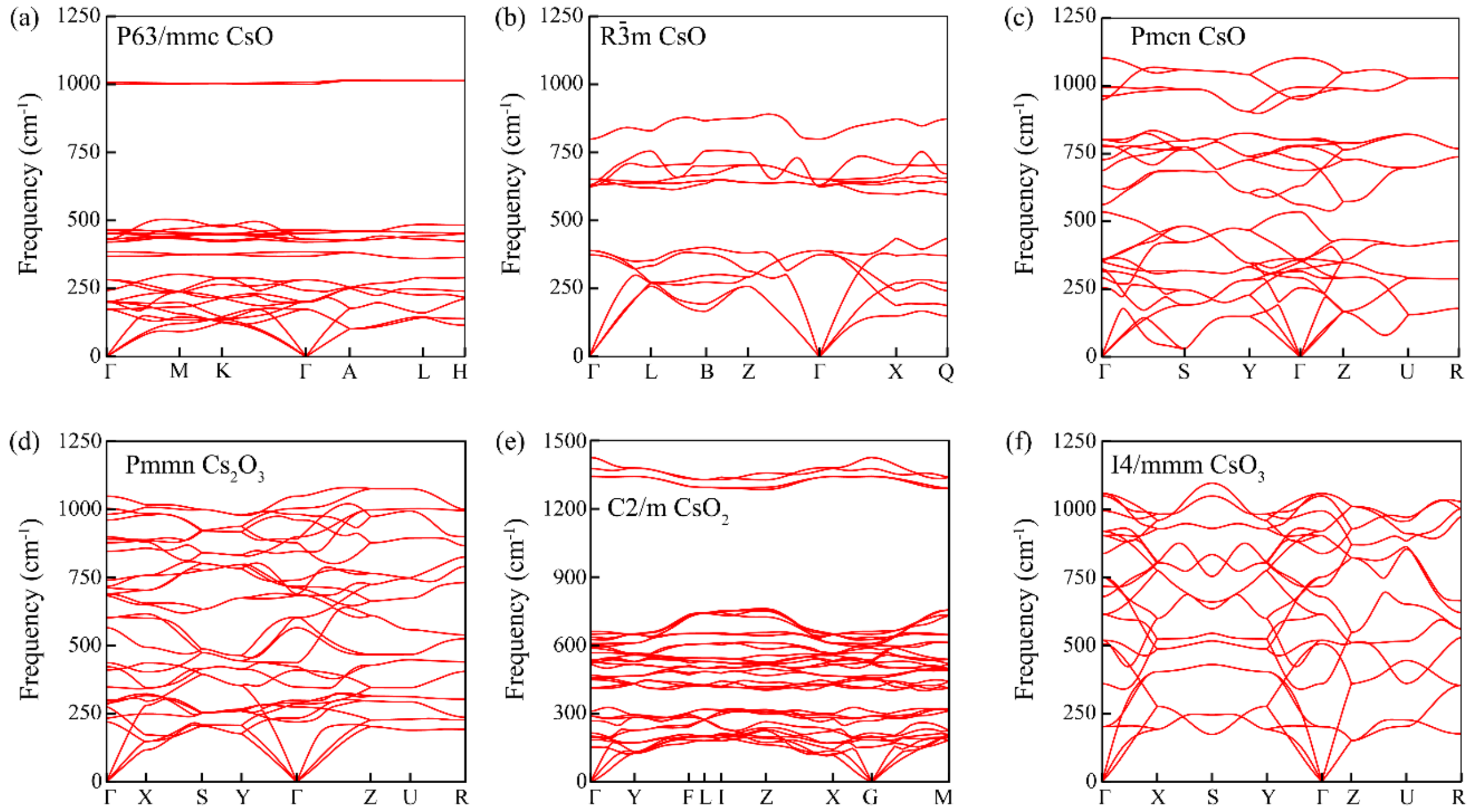

**Figure S2.** Phonon spectrum of (a) CsO in $P6_3/mmc$ symmetry at 100 GPa. (b) CsO in $R\bar{3}m$ symmetry at 300 GPa. (c) CsO in *Pmcn* symmetry at 500 GPa. (d) $Cs_2O_3$ in *Pmmn* symmetry at 400 GPa. (e) $CsO_2$ in *C2/m* symmetry at 200 GPa. (f) $CsO_3$ in *I4/mmm* symmetry at 400 GPa.

### 3. Electronic properties of Cs-O compounds

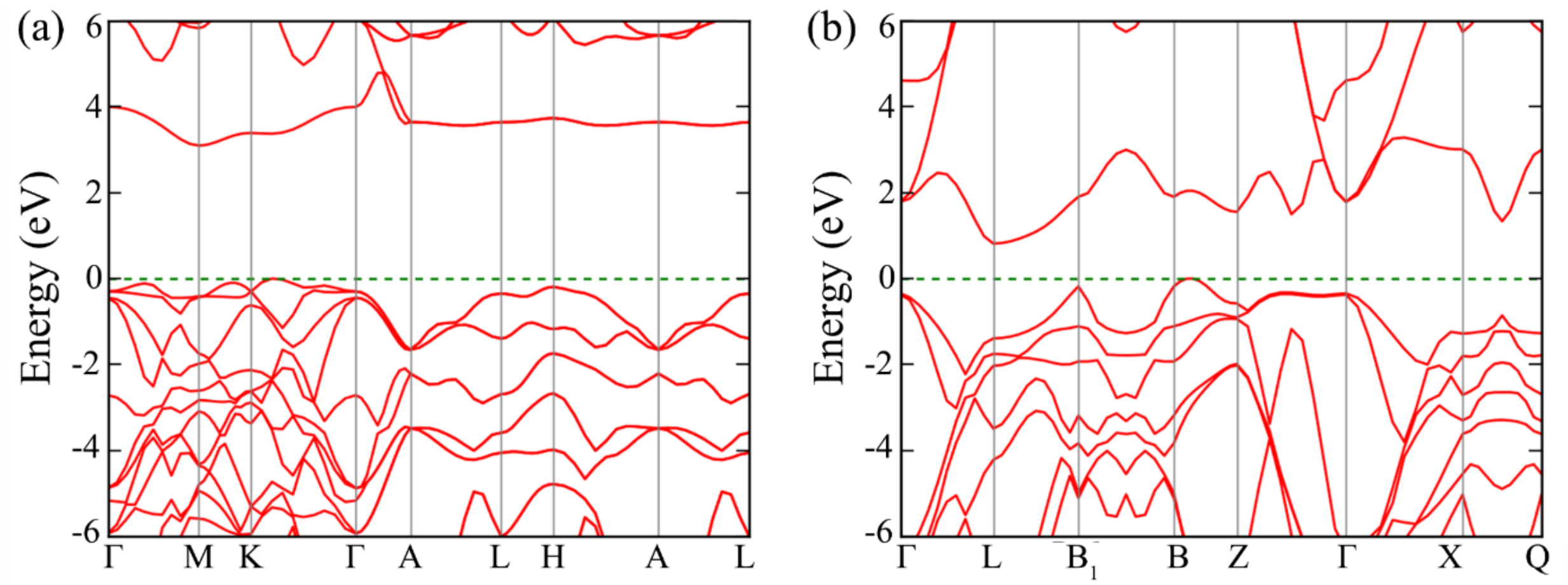

**Figure S3.** Band structures of (a) *P6₃/mmc* CsO and (b) $R\overline{3}m$ CsO at 221 GPa, respectively.

## 4. Bonding analysis of Cs-O compounds

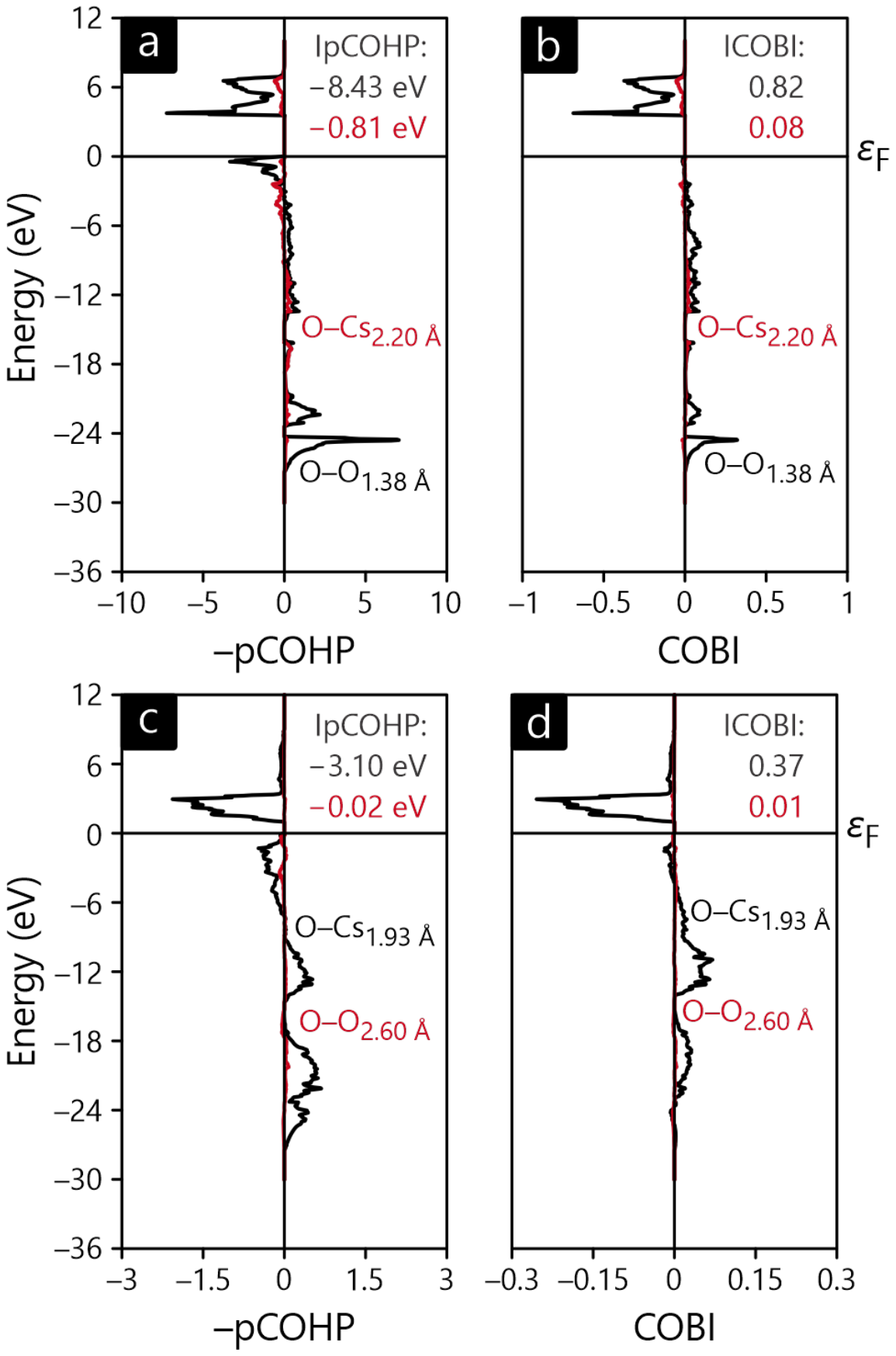

**Figure S4.** Integrated crystal orbital Hamiltonian population (COHP, left) and crystal orbital bond index (COBI, right) of (a) *P6₃/mmc* CsO and (b) $R\bar{3}m$ CsO under 221 GPa, respectively.

## 5. Experimental methods to identify Cs-O compounds

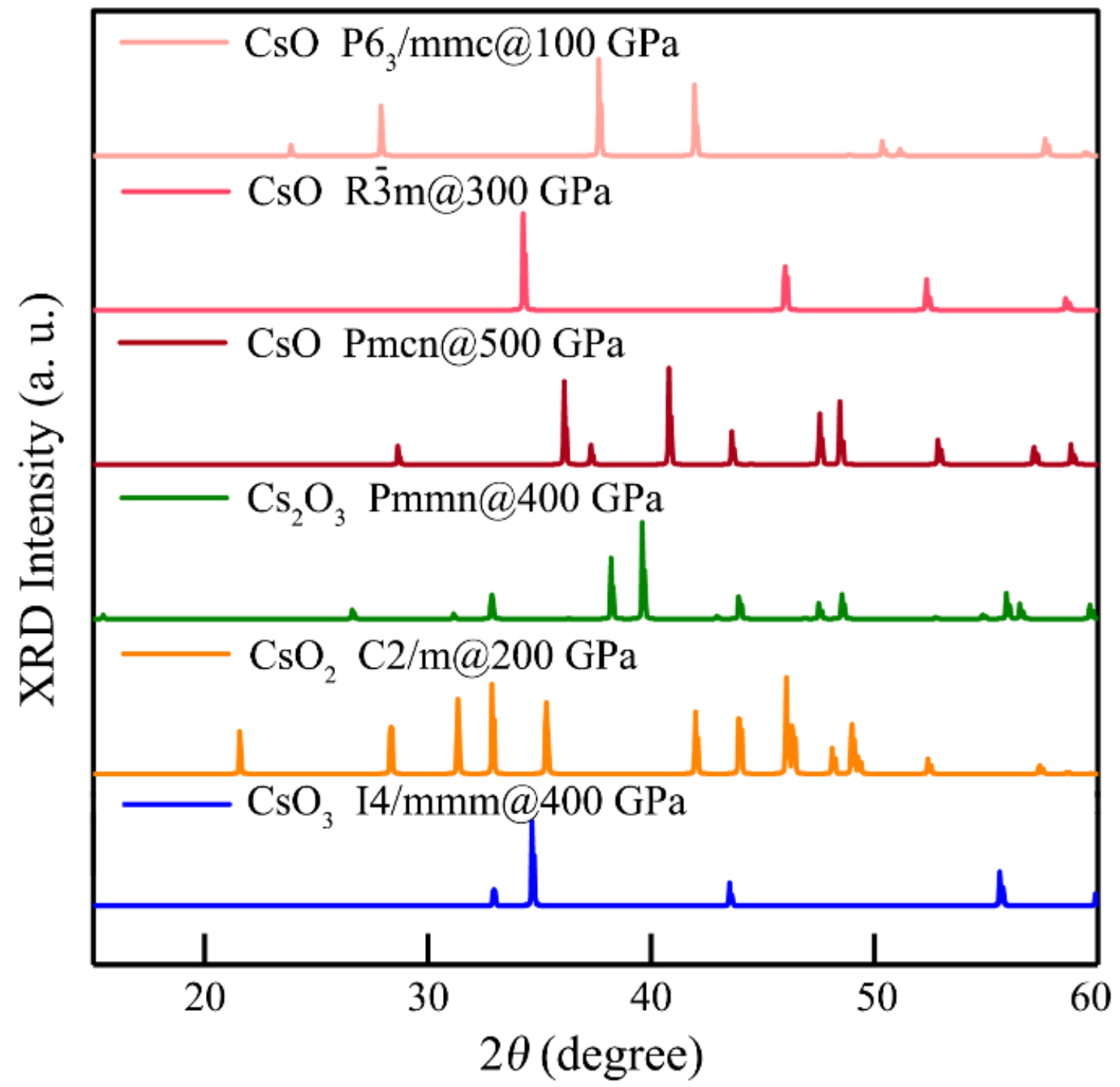

**Figure S5.** The simulated XRD patterns of CsO in *P6₃/mmc* symmetry, CsO in $R\bar{3}m$ symmetry, CsO in *Pmcn* symmetry, $Cs_2O_3$ in *Pmmn* symmetry, $CsO_2$ in *C2/m* symmetry, and $CsO_3$ in *I4/mmm* symmetry.

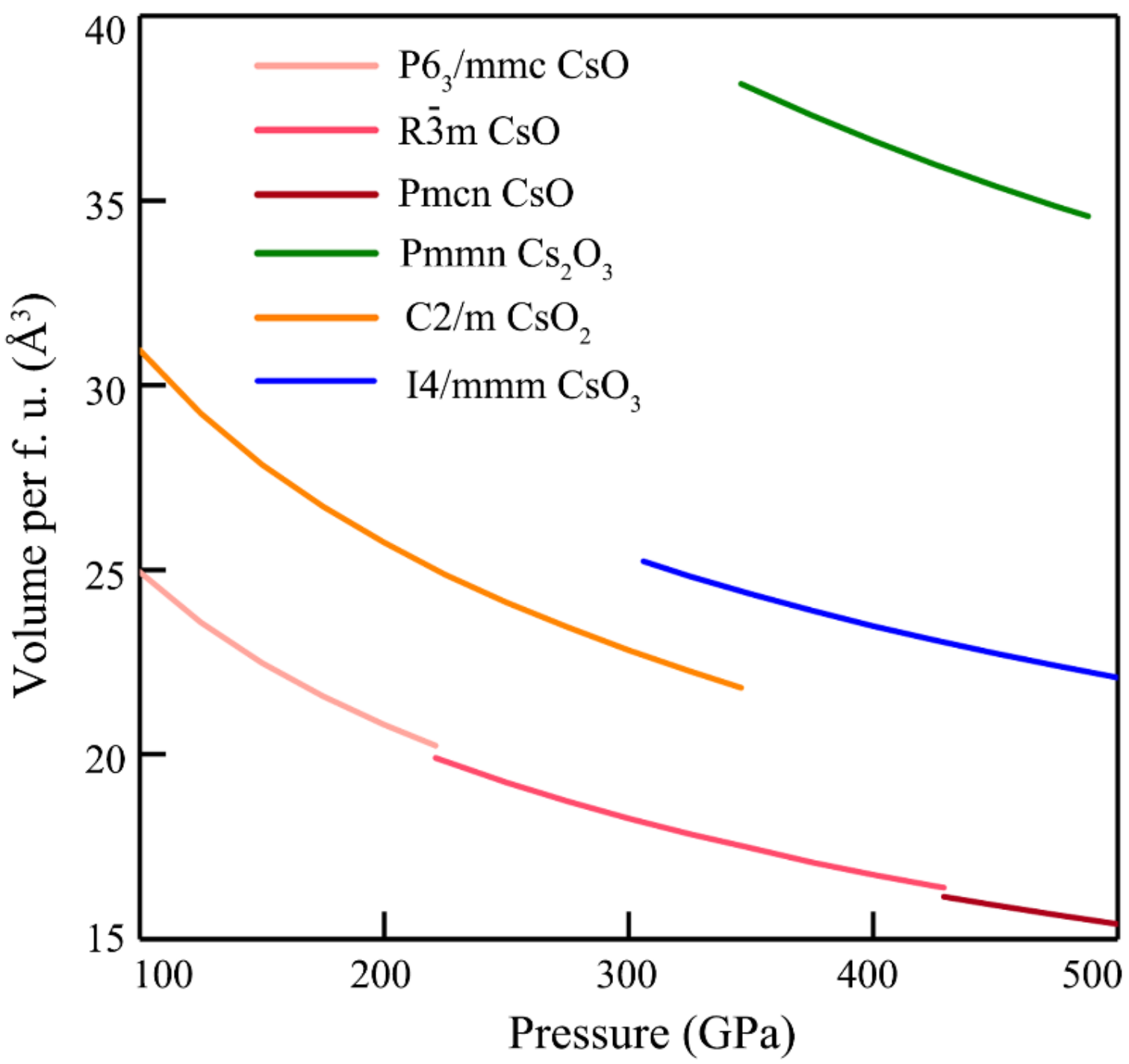

**Figure S6.** The simulated equation of states of CsO in *P6₃/mmc* symmetry, CsO in $R\bar{3}m$ symmetry, CsO in *Pmcn* symmetry, $Cs_2O_3$ in *Pmmn* symmetry, $CsO_2$ in *C2/m* symmetry, and $CsO_3$ in *I4/mmm* symmetry.

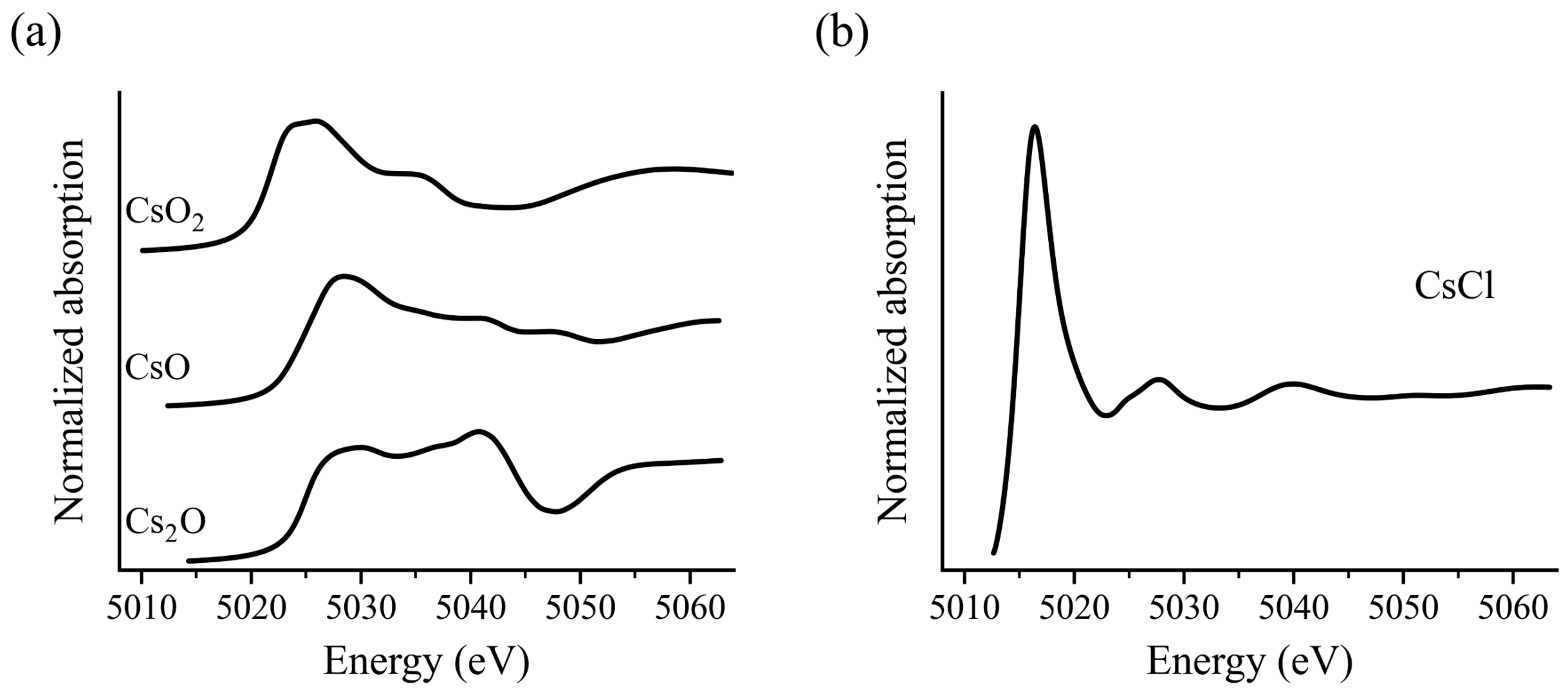

**Figure S7.** Calculated Cs $L_3$-edge XANES spectra of (a) $Cs_2O$, CsO, and $CsO_2$ at 250 GPa and (b) CsCl at ambient pressure. The spectra were normalized in the plotted energy range, and no horizontal energy shift was applied. The calculated CsCl spectrum agrees well with the experimental CsCl Cs $L_3$-edge XANES reference reported previously, including the sharp white-line feature and weak post-edge structure.[23,24]

## 6. Structural information of stable Cs-O compounds

**Table S1.** Calculated structural parameters of various stable Cs-O compounds.

| | Space group | Lattice parameters (Å, °) | Atomic coordinates (fractional) | | | |
|---|---|---|---|---|---|---|
| | | | Atoms | X | Y | Z |
| CsO (100 GPa) | $P6_3/mmc$ | a = b = 4.3029<br>c = 6.2159<br>$\alpha = \beta$ = 90.0000<br>$\gamma$ = 120.0000 | Cs<br>Cs<br>O | 0.6667<br>0.0000<br>0.3333 | 0.3333<br>0.0000<br>0.6667 | 0.7500<br>0.0000<br>0.8652 |
| CsO (300 GPa) | $R\bar{3}m$ | a = b = c = 3.5399<br>$\alpha = \beta = \gamma$ = 67.7178 | Cs<br>Cs<br>O | 0.0000<br>0.5000<br>0.2356 | 0.0000<br>0.5000<br>0.2356 | 0.0000<br>0.5000<br>0.2336 |
| CsO (500 GPa) | *Pmcn* | a = 3.1378<br>b = 4.0764<br>c = 4.8192<br>$\alpha = \beta = \gamma$ = 90.0000 | Cs<br>O | 0.2500<br>0.7500 | 0.8677<br>0.6385 | 0.6772<br>0.4656 |
| $Cs_2O_3$ (400 GPa) | *Pmmn* | a = 5.7363<br>b = 4.1196<br>c = 3.0980<br>$\alpha = \beta = \gamma$ = 90.0000 | Cs<br>Cs<br>O<br>O | 0.3576<br>0.0491<br>0.2787<br>0.5828 | 0.5000<br>0.0000<br>0.7671<br>0.5000 | 0.0000<br>0.0000<br>0.5000<br>0.5000 |
| $CsO_2$ (200 GPa) | *C2/m* | a = 6.6226<br>b = 5.4464<br>c = 3.0003<br>$\alpha = \gamma$ = 90.0000<br>$\beta$ = 108.0569 | Cs<br>O | 0.6984<br>0.9409 | 0.0000<br>0.1805 | 0.9446<br>0.6328 |
| $CsO_3$ (400 GPa) | *I4/mmm* | a = b = 2.9398<br>c = 5.4361<br>$\alpha = \beta = \gamma$ = 90.0000 | Cs<br>O<br>O | 0.5000<br>0.0000<br>0.5000 | 0.5000<br>0.0000<br>0.0000 | 0.5000<br>0.5000<br>0.7500 |

## 7. Analysis of band characters of Cs-O compounds

**Table S2.** ICOHPs and ICOBIs for the nearest neighboring Cs-Cs, Cs-O, and O-O pairs for Cs-O compounds under pressure.

| | **Space group** | **Atomic pairs** | **Distance (Å)** | **ICOHP (eV/pair)** | **ICOBI** |
|---|---|---|---|---|---|
| CsO (221 GPa) | $P6_3/mmc$ | Cs-Cs | 2.734 | -1.393 | 0.105 |
| | | Cs-O | 2.202 | -0.806 | 0.077 |
| | | O-O | 1.390 | -8.426 | 0.826 |
| CsO (221 GPa) | $R\bar{3}m$ | Cs-Cs | 2.729 | -1.426 | 0.104 |
| | | $Cs^1$-O | 1.940 | -3.097 | 0.368 |
| | | $Cs^2$-O | 2.211 | -0.719 | 0.064 |
| | | O-O | 2.603 | -0.017 | 0.007 |
| CsO (500 GPa) | *Pmcn* | Cs-Cs | 2.558 | -1.208 | 0.083 |
| | | Cs-O | 2.046 | -1.030 | 0.075 |
| | | O-O | 1.962 | -0.070 | 0.026 |
| $Cs_2O_3$ (400 GPa) | *Pmmn* | Cs-Cs | 2.638 | -1.161 | 0.088 |
| | | $Cs^1$-O | 1.953 | -2.376 | 0.228 |
| | | $Cs^2$-O | 2.111 | -0.486 | 0.028 |
| | | O-O | 1.919 | -0.194 | 0.016 |
| $CsO_2$ (200 GPa) | *C2/m* | Cs-Cs | 2.749 | -1.821 | 0.133 |
| | | Cs-O | 2.314 | -0.437 | 0.043 |
| | | O-O | 1.278 | -13.064 | 1.037 |
| $CsO_3$ (400 GPa) | *I4/mmm* | Cs-Cs | 2.926 | -1.540 | 0.102 |
| | | Cs-O | 2.002 | -2.051 | 0.183 |
| | | O-O | 2.002 | -0.159 | 0.031 |